# No compelling evidence for clathrate hydrate formation under interstellar medium conditions over laboratory time scales

Mathieu Choukroun[a,1], Tuan H. Vu[a], and Edith C. Fayolle[a]

In PNAS, Ghosh et al. (1) report their experimental observations of methane and $CO_2$ clathrate formation at conditions similar to the interstellar medium (ISM), namely 10 to 30 K and $10^{-10}$ mbar. The authors conducted time-dependent reflection–absorption infrared spectroscopy (RAIRS) of vapor-deposited $H_2O$–$CH_4$ and $H_2O$–$CO_2$ mixtures and interpreted new blue and red shifted peaks from those of trapped $CH_4$ and $CO_2$ in amorphous ice, respectively, as indicative of clathrate formation. In this letter, we point out potential pitfalls and caution against the implications drawn for the ISM.

In the case of the $H_2O$–$CO_2$ system, Ghosh et al. (1) attribute the observed red shift in the infrared (IR) spectra to hydrogen bonding between $CO_2$ and the water cage. This statement not only goes against the common knowledge of $CO_2$'s lack of affinity for hydrogen bonding, but also contradicts the calculated structure in figure 2 of ref. 1, which clearly shows that there is no available hydrogen atom in the $5^{12}$ cage for the $CO_2$ guest to form hydrogen bonds with. Any hydrogen bonding with the skeleton would greatly distort the cage structure, as has been suggested for $NH_3$-bearing clathrates (2). Furthermore, the O–H stretching band in figure S4 of ref. 1 is characteristic of amorphous ice (3), while clathrate hydrates are crystalline compounds.

Ghosh et al. (1) argue that the close match between experimentally observed shifts and computed shifts indicates formation of a clathrate $5^{12}$ cage. However, table S1 of ref. 1 claims a shift of $-36.0$ $cm^{-1}$ for the $CO_2$ antisymmetric stretch (although figure 2 ref. 1 shows a shift of $-7$ $cm^{-1}$), whereas the computed shift is $-86.0$ $cm^{-1}$.

The emergence of a 2,346 $cm^{-1}$ $CO_2$ peak is interpreted as due to clathrate formation by analogy to an earlier study (4). However, in that work, the chemical system also contained $CH_3OH$, whose presence appears necessary for $CO_2$ clathrates to form after slow heating to 120 K. In addition, $CO_2$ clathrates grown epitaxially on other clathrates (5) or formed under higher pressure/temperature conditions (6) have a characteristic IR signature with double peaks consistently observed in the $\nu_3$ regions of both $^{12}CO_2$ and $^{13}CO_2$, unlike the signatures observed in figure 2 and figure S5 of ref. 1. Furthermore, RAIRS studies of layered and mixed $H_2O$–$CO_2$ ices (7, 8) reported a peak around 2,346 $cm^{-1}$, which was interpreted in these studies as $CO_2$ interacting with or being trapped within amorphous $H_2O$ [albeit not as forming a clathrate as inferred by Ghosh et al. (1)].

For the $H_2O$–$CH_4$ system, the calculated blue shift for methane in the $5^{12}$ cage (1) is inconsistent with previous experimental studies showing that enclathrated methane exhibits a red shift (9, 10). Predicted red shifts for other cage types and for $CO_2$ presented in the present study (1) are also consistent with this general behavior.

In summary, we find that the experimental data and theoretical computations in Ghosh et al. (1) do not support the interpretation of clathrate hydrate formation. The community should be cautioned against interpreting this article as definitive evidence that clathrates can be formed in the ISM within short (approximately days) laboratory time scales.

1 J. Ghosh *et al.*, Clathrate hydrates in interstellar environment. *Proc. Natl. Acad. Sci. U.S.A.* **116**, 1526–1531 (2019).
2 K. Shin, R. Kumar, K. A. Udachin, S. Alavi, J. A. Ripmeester, Ammonia clathrate hydrates as new solid phases for Titan, Enceladus, and other planetary systems. *Proc. Natl. Acad. Sci. U.S.A.* **109**, 14785–14790 (2012).
3 W. Hagen, A. Tielens, J. Greenberg, The infrared spectra of amorphous solid water and ice Ic between 10 and 140 K. *Chem. Phys.* **56**, 367–379 (1981).

[a]Jet Propulsion Laboratory, California Institute of Technology, Pasadena, CA 91109
Author contributions: M.C., T.H.V., and E.C.F. wrote the paper.
The authors declare no conflict of interest.
Published under the PNAS license.
[1]To whom correspondence may be addressed. Email: mathieu.choukroun@jpl.nasa.gov.